# Flat-band energy analysis of the superconducting gap for hydrogenated graphite fibers found from nonlocal electrical conductance experimental data

Nadina Gheorghiu, Charles R. Ebbing, and Timothy J. Haugan

*Abstract*—**Experimental evidence of novel phenomena in hydrogenated graphite fibers is found. An indirect excitonic mechanism is likely leading to a SC state below the temperature $T_c \cong 50$ K, where the gap is divergent. Analysis of the gap within the framework provided by the Bardeen-Cooper-Schrieffer (BCS) theory of superconductivity shows that this is a multigap system. The energy gap data can be better explained within the framework of topologically protected flat bands applied to systems in which superconductivity occurs on the surface or at the internal interfaces of the samples. The temperature dependence of the SC gap is linear above 50 K. We use nonlocal differential conductance $G_{diff}(V) = dI(V)/dV$ experimental data to show clear evidence of topological phenomena such as interference of chiral asymmetric Andreev edge states and crossed Andreev conversion. $G_{diff}(V)$ has a negative part that results from the nonlocal coherence between electron and holes in the Andreev edge states. We conclude that hydrogenated graphite bears the marks of an unconventional high-temperature superconductor (HTSC).**

## I. Introduction

OPTIMAL electronic systems require dissipationless circuits. Superconducting (SC) materials are the obvious choice. At the same time, applications like spintronics require magnetic materials. Twisted few-layer graphene shows both magnetism and superconductivity, which were for long considered mutually excluded phenomena. While graphene is the model system for any novel two-dimensional (2D) material, its 3D parent system – graphite - can reveal similar and other unexpected quantum phenomena. We have previously reported on HTSC in graphite fibers and other carbon(C)-based systems [1]–[4].

While $G_{diff}(V)$ measurements for SC systems are usually done locally using scanning tunneling spectroscopy, a highly disordered system like graphite fibers needs nonlocal $G_{diff}(V)$ measurements. Unlike the former, nonlocal $G_{diff}(V)$ data can distinguish between nontopological zero-energy modes that are localized around potential inhomogeneities and true Majorana edge-bound modes of the topological phase [5]. As known, the Majorana effect is fundamental for the function of fault-tolerant quantum computers. Considering normal metal leads connected to a SC, two processes are at the core of nonlocal response in $G_{diff}(V)$: 1) direct electron transfer between the normal leads; 2) crossed Andreev reflection [6] (an incoming electron is converted into a reflected hole) of an electron from one lead into another lead. The height of the peak in $G_{diff}(V)$ is proportional to the nonlocal density of states (DOS), while the energy gap, defined as the width of $G_{diff}(V)$ at half-height, is a measure of correlations between electrons potentially leading to the formation of Cooper pairs.

An important class of HTSC materials are the C fiber-HTSC hybrids that are used as stronger, flexible, and chemically stable HTSC wires for SC magnets, particle accelerators, nuclear magnetic resonance (NMR) devices or electromagnetic interference shielding covers for spacecrafts. Quantitively, NbN-coated C fibers have critical densities $J_c \sim 10^6$ A/cm$^2$ [7] and critical fields $B_{c2}(0)$ up to 25 T [8].

The physical processes behind the observed temperature($T$)-dependent transport properties of disordered systems such as C fibers have been reviewed [9]. The disorder forces the electrons to confine within the lattice, a phenomenon known as localization. The random potential, which quantifies the effects of disorder, directs the localization of electrons within the region of the trap. Energy optimization causes the hopping of an electron to an energy level close to the one for a neighboring state. The trapping results in inelastic scattering in the electron-electron (*e-e*) interactions. The *T*-dependent electrical conductivity is described by Mott law: $\sigma_T = \sigma_0 \exp[-(T_M/T)^{1/(d+1)}]$, where $T_M$ is the Mott temperature and the power factor $d = 0, 1, 2,$ or $3$ quantifies the dimensionality of the electronic transport [10]. Of interest here, the trapped electrons can interact and form Cooper pairs. In SC materials, disorder can play an intricate role. Trapped charges can suppress the coherence of phase slips, thus favoring SC correlations [11]. An

This work was supported by The Air Force Office of Scientific Research (AFOSR) for the LRIR #14RQ08COR & LRIR #18RQCOR100 and the Aerospace Systems Directorate (AFRL/RQ). We acknowledge J.W. Lawson for the SEM, J.P. Murphy for the cryogenics, and Dr. T.J. Bullard for discussions. N. Gheorghiu acknowledges Dr. G.Y. Panasyuk for his continuous support and inspiration.

*Corresponding author: N. Gheorghiu*

N. Gheorghiu was with UES, Inc., Dayton, OH 45432, USA. (Nadina.Gheorghiu@yahoo.com).
C.R. Ebbing is with University of Dayton Research Institute, Dayton, OH 45469, USA.
T.J. Haugan is with The Air Force Research Laboratory (AFRL), The Aerospace Systems Directorate, AFRL/RQ, Wright-Patterson AFB, OH 45433, USA.

increase in the 2D density of charges through material intercalation can enhance the likelihood for the occurrence of quantum phenomena such as 2D SC [12].

## II. EXPERIMENT

SC has been often found in the unsuspected materials. Here, the choice is for polyacrylonitrile $((CH_2\text{-}CH\text{-}CN)_n)$ T300 fibers having the average diameter 7 μm. Intercalation with 99.99% purity octane ($C_8H_{18}$) results in the formation of hydrogen(H)-rich puddles between neighboring graphite layers. The new system is a H-C-N fiber. While we have relied on the Van der Pauw method [13] for disc- or square-shape area samples, here the four-point contact method was used. Four ~2 mm wide silver (Ag) strips were affixed to a sapphire substrate via C dots. The H-C-N fiber was transversely placed and affixed on the Ag strips by dot-like colloidal Ag contacts. The current gap was 11 mm, while the voltage gap was from 2.1 mm. For a good thermal contact, cryogenic grease was applied between the fiber and the substrate. The quality of the electrical contacts was optically checked using an Olympus BX51 microscope equipped with a digital camera (Fig. 1a). A 3D view of the fiber was obtained using scanning electron microscopy (Fig. 1b). The sample was placed on an aluminum heater block and four POGO pins spring-pressed on the H-C-N fiber completed the measurement circuit. Transport measurements without magnetic field were carried out with the sample in a Gifford-McMahon cryocooler and the temperature set by a LakeShore 340 Controller. Direct electrical current $I$ was sourced through a Keithley 2430 1 kW PULSE current-source meter and the voltage $V$ along the sample was measured with a Keithley 2183A Nanovoltmeter.

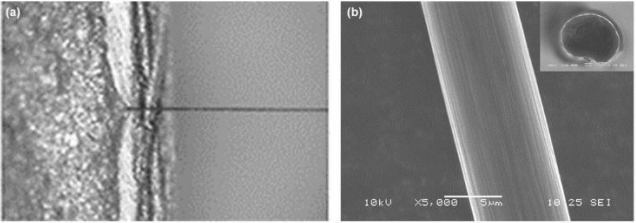

Fig. 1. Microscopy of a H-C-N fiber: a) Optical microscopy showing one end of the fiber firmly connected to a silver contact. b) Scanning electron microscopy.

## III. RESULTS AND DISCUSSION

Nonlocal $G_{diff}(V)$ experimental data as a function of the applied voltage $V$ is shown in Fig. 2. Increased density of free charges results after increasing the sourced current. One remarkable feature is the asymmetry in $G_{diff}(V)$. Separated in space and bound by the Coulomb (attractive) interaction, the electrons and holes 'crystallize" into particle(electron)-hole $e-h$ pairs pretty much like electric dipoles. When the $e-h$ symmetry and the time reversal symmetries are violated, the differential tunneling (local) electrical conductivity and the dynamic (nonlocal) $G_{diff}(V)$ are no more symmetric functions of $V$ [14]. This asymmetry can be observed both in the normal and SC phases of strongly correlated systems. As in normal Fermi liquids the $e-h$ symmetry is not violated, the differential tunneling conductivity and the dynamic conductance are symmetric functions of $V$. Thus, asymmetric conductance is not observed in conventional metals, especially at low T. In fact, the asymmetry in the tunneling conductance is an important sign of the underlying Mott character in doped insulating systems [15] that can show unconventional SC.

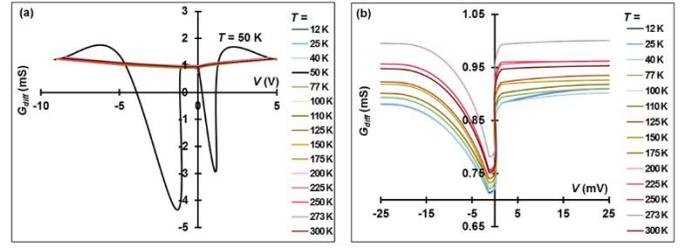

Fig. 2. Nonlocal $G_{diff}(V)$ data for an octane-intercalated H-C-N fiber at high $V$ (a) and low $V$ (b), respectively.

The energy gap $\Delta$ is determined as shown in Fig. 3a. Significantly, the non-zero $G_{diff}$ at almost zero-$V$ gap suggests the possible existence of Majorana fermions edge modes. Another remarkable feature seen in Fig. 2a is the shape of $G_{diff}(V)$ for $T = 50$ K. In this case, the asymmetry in $G_{diff}(V)$ (Fig. 3b) mirrors the interference of chiral Andreev edge states, which is a topological phenomenon. In addition, negative $G_{diff}(V)$ is a result of nonlocal coherence between electrons and holes in the Andreev edge states. Thus, negative $G_{diff}(V)$ indicates the presence of Andreev converted holes, namely the coupling between two quantum Hall edge states via a narrow SC link. The inset in Fig. 3b shows that the hump spreads from $-\Delta$ to $\Delta$: $2\Delta \cong (1.34 + 1.11)$ eV $\cong 2.5$ eV. This is about 400× the hump width for YBCO. We are dealing with a chiral spin-triplet $p$-wave with two gap amplitudes: $\Delta_{V<0} \cong 1.6$ eV for the hump located at $V \cong -1.34$ V and $\Delta_{V>0} \cong 0.6$ eV for the hump located at $V \cong 1.11$ V, respectively. Notice that the two minima are located at voltages of the order of $\Delta/|e|$ [16]. Interestingly, we have found another minimum for $G_{diff}(V)$ for $T = 130$ K corresponding to a voltage $V \cong 63$ meV (Fig. 10 in [2]). This would correspond to a $T_c = \Delta/k_B \cong 350$ K. Significantly, the corresponding frequencies $f = \Delta/h$ are extremely high, of about 380 THz and 145 THz, respectively. While graphene or C nanotubes have proven imaging and sensing capabilities in the microwave domain (such as the use of bolometers for detecting the electronic specific heat), a H-C-N fiber seemingly can detect resonances in the THz range of the electromagnetic spectrum, showing potential for ultra-fast read out applications, such as in communications. At the same time, the two minima in $G_{diff}(V)$ are evidence of topological insulating states realized by the formation of two internal excitons, i.e., $e-h$ quasiparticle condensations. This finding is important, as Majorana bound states are predicted to occur in a spin-triplet SC. For $V$ smaller than the SC gap at this temperature (50 K), the transport is dominated by the Andreev reflection. The gap asymmetry is due to charge imbalance that creates different rates at which the $e$–like and $h$–like quasiparticles are evacuated from the Andreev bound states and/or by the destruction of chirality symmetry by the magnetic exchange field due to the itinerant ferromagnetism (FM) introduced in the system by octane with its freely moving H+ (protons) on the interfaces of graphite [17] and HTSC graphite fibers [18]. We also notice the Fano-line shape of $G_{diff}(V)$. The Fano line is a manifestation of coexisting polarons and Fermi particles in a superlattice of quantum wires. Fano resonances can lead to multi-gap HTSC, as we will see later. In the stripe scenario for HTSC, a Fermi liquid coexists with an incommensurate 1D charge density wave forming a multi-gap SC near a Lifshitz transition where the critical $T_c$ amplification is driven by Fano resonances involving different condensates. Thus, $G_{diff}(V)$ data shows that SC correlations might be established in the H-C-N fibers below $T_c \sim 50$ K. This value is close to the mean-field $T$ for SC correlations in the metallic-H multilayer graphene or in highly oriented pyrolytic graphite, $T_c \sim 60$ K [19], [4].

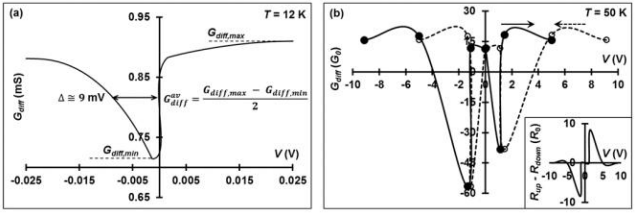

Fig. 3. a) Example calculation of the SC gap for the H-C-N fiber at $T = 12$ K. b) The gap at $T = 50$ K suggests interference of chiral asymmetric Andreev edge states and crossed Andreev conversion.

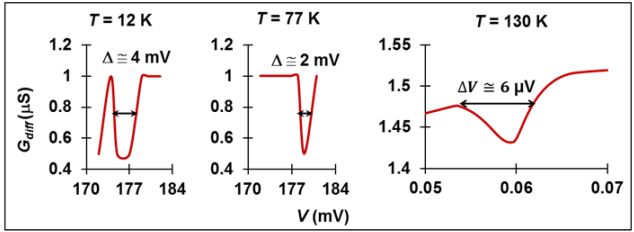

Fig. 4. Small gap data at non-zero $V$ at $T = 12$ K, $T = 77$ K, and $T = 130$ K, respectively.

We have also observed gaps in the $G_{diff}(V)$ data at other small, nonzero $V$ (Fig. 4). Fitting the large gap and the small gap data to the BCS formula $\Delta(T) = \Delta(0) \sqrt{1 - (T/T_c)^2}$ [20], we find that the H-C-N fiber is a multiple-gap system (Fig. 5). For $T \leq 130$ K, both a small gap $\Delta_S(0) \cong 6.5$ meV and a large $\Delta_L(0) \cong 13$ meV gap are observed. At $T_c \cong 50$ K, the gap ratios are: $2\Delta_S(0)/k_B T_c \cong 3.0$ and $2\Delta_L(0)/k_B T_c \cong 6.0$, respectively. Interestingly, these are close to the universality class values found for MgB$_2$ [21] and for other SC materials [22]. For comparison, the phonon-mediated SC MgB$_2$, which is crystallographically and electronically similar to graphite, has a small gap $\Delta_S$ = 4.4 meV and a large gap $\Delta_L$ = 9.6 meV gap. The $r_{BCS} = \Delta(0)/k_B T_c$ ratios (TABLE I), where $k_B = 1.38 \times 10^{-23}$ J/K is the Boltzmann constant, are smaller than the known BCS value of 1.74, suggesting unconventional HTSC.

Interestingly, the small gap $\Delta_S(T)$ data is almost a perfect line (black line in Fig. 5), intersecting the $T$-axis at $T_c \cong 130$ K. The linear $\Delta_S(T)$ dependence suggests the spin Seebeck effect due to the spin-orbit interaction, where the gap $\Delta_S$ is a measure of the magnetic resonance energy $E_r \propto \Delta_S$ [23]. It also points to the excitonic mechanism for unconventional SC based on the existence of a spin exciton, i.e. a spin-1 $e$–$h$ excitation. The excitonic mechanism [24], [25] can lead to HTSC, with possible $T_c \sim 2200$ K in organic systems [26]. Importantly, the spin-orbit coupling can lead to localized Majorana edge modes in this ferromagnetic SC (FMSC).

The BCS fit to all gap data below 300 K intersects at $T_c \cong 440$ K. In the BCS theory, $T_c = T_D e^{-1/[N(0)V_{e-e}]}$, where $g = N(0)V_{e-e}$ is the electron-phonon coupling constant, with $N(0)$ the density of states near the Fermi surface in the normal state and $V_{e-e}$ the matrix element of the attractive potential between electrons. Thus, the high coupling constant $g \cong 0.62$ for H-C-N fiber suggests that the strength of $e$–$e$ correlations is larger than the one expected via phonon coupling. On the other hand, a Debye temperature lower than the one known for graphite, $T_D < 2280$ K (the known value for graphite fibers [27]), would preserve the otherwise known Ginzburg factor $g = 0.5$. A $T_D \cong 725$ K would be also consistent with the metallic nature of the H-C-N fiber that is due to the free protonation of graphite interfaces. At the same time, a $T_D \cong 725$ K comes close to both an estimated $T_c = 2\Delta/(3.5k_B) \cong (2 \times 63 \text{ meV})/(3.5k_B) \cong 730$ K corresponding to the minimum in $G_{diff}(V)$ for higher $V$ and at $T = 130$ K (see the note before) and, though less close, to the value $T_c \cong 576$ K resulted from fitting experimental data $\Delta(T)$ to the BCS formula (TABLE I).

TABLE I
$\Delta(0)$ AND $T_C$ DETERMINED FROM FITTING
$\Delta_{EXP}(T)$ TO $\Delta_{BCS}(T) = \Delta(0) \sqrt{1 - (T/T_c)^2}$

| $T$ range (K) | $\Delta(0)$(meV) | $T_c$(K) | $r_{BCS}$ |
| --- | --- | --- | --- |
| $T < 50$ K | 14.4 | 120 | 1.4 |
| $50$ K $< T < 140$ K | 15.5 | 179 | 1.0 |
| $140$ K $< T < 300$ K | 11.9 | 576 | 0.24 |
| $50$ K $< T < 300$ K | 12.6 | 475 | 0.31 |
| $T < 140$ K | 14.3 | 208 | 0.80 |
| $T < 300$ K | 13.0 | 440 | 0.36 |

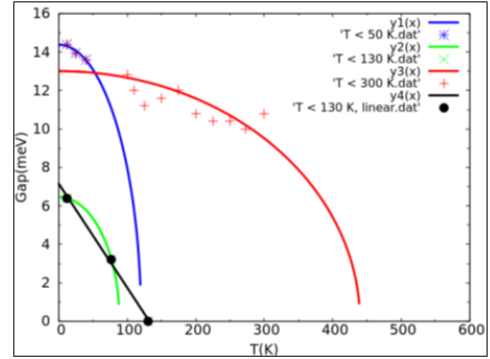

Fig. 5. Experimental $\Delta(T)$ and the fit to the BCS formula. The large gap data is for $T$ below 130 K (in blue) and $T$ up to 300 K (in red). The small gap data (fitting curve in green) is better fit by a line (in black).

The $\Delta(0)$ vs. $T_c$ plot is shown in Fig. 6a. A linear fit of the data gives (a zero gap) at a $T_c \cong 2360$ K. This result is in accordance with Schrieffer's prediction for exotic HTSC, which states that a $p$-pairing for FM spin fluctuations would lead to a very large $T_c$ [28]. Interestingly, within the two-band gap model used for MgB$_2$, the BCS-like gap $\Delta(0)$ increases linearly with $T_c$, while $\Delta(0)$ characteristic to organic SCs and alkane-graphite systems decreases linearly with $T_c$ [29].

The ratio $2\Delta(0)/k_B T_c$ was also found to increase with the coefficient Γ defined within a set of quantum-critical models in which the pairing interaction is mediated by a gapless boson with local susceptibility given in the Γ model by $\chi(\Omega) = 1/|\Omega|^\Gamma$ [30], [31], with Ω as the renormalized phonon frequency. The ratio $2\Delta(0)/k_BT_c$ has been recently computed numerically for $0 < \Gamma < 2$ within Eliashberg theory and was found to increase with increasing Γ. It was argued that the origin of the increase is the divergence of $2\Delta(0)/k_BT_c$ at $\Gamma = 3$. At the divergence, the gap was found to scale with $T_c$ as: $2\Delta(0)/T_c \approx 4\pi(1-\Gamma)^{-1/3}$. Indeed, the gap divergent value for the H-C-N fiber $\Delta(T = 50$ K$) = 1.6$ eV gives $\Gamma \cong 3$ (Fig. 6b). Moreover, the ratio $2\Delta(0)/k_BT_c$ at the divergence is about 8, a value that is characteristic to HTSC materials [32].

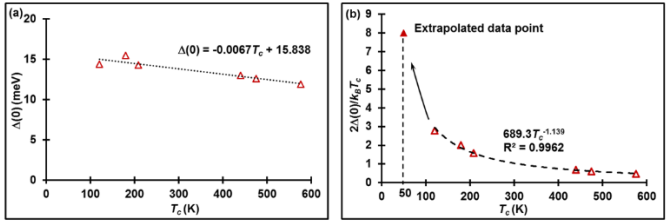

Fig. 6. a) The gap $\Delta(0)$ dependence on $T_c$. Data from TABLE I. b) Power law dependence of the BCS ratio as a function of $T_c$, showing the divergence at $T_c \cong 50$ K, where the BCS ratio is $r_{BCS} \cong 8$.

Perhaps the best way to analyze the gap data for graphite-based materials is within the framework of topologically protected zero-energy bands or flat-energy bands (FB). It is suggested that in order to reach room-$T$ SC, one must search for or artificially create systems that experience nontopological FB in the bulk or topologically protected FB on the surface or at the interfaces of the samples [33], [34]. The evidence herein presented shows that FBs might occur in these disordered hydrogenated graphite fibers, which have their nm-size domains of parallel C layers twisted with respect to each other. This structure is similar to helical structures of disc-shaped molecules of liquid crystals where the twist is provided by the intermolecular H bonds. As known, the twist distortion that played an important role in the the analogy between SCs and smectic A liquid crystals lead to better understanding of the former [35]. The topological origin of the FB can be also understood in terms of the pseudo-magnetic field created by the internal strain and results in FM behavior [36]. As the FB states are highly localized around certain spots in the structure, the SC order parameter becomes strongly inhomogeneous. The $\Delta(T)$ dependence is the solution to a transcendental equation: $\Delta(T) = \Delta_{FB}\tanh[\Delta(T)/(2k_BT)]$. This $\Delta(T)$ dependence comes from the density functional theory approach for SC, where the extended Kohn-Sham equation is written in the form of Bogoliubov-deGennes equation used in the conventional theory for the description of inhomogeneous SC. At the transition, the condition for the FB energy gap $\Delta_{FB}$ translates as $\Delta(T_c) \rightarrow 0$. Taylor series expansion to the first order gives $T_c \simeq \Delta_{FB}/(2k_B)$. Our results for the H-C-N fiber are shown in Fig. 7. The $\Delta(T)$ data for the H-C-N fiber was replotted with the constant $\Delta_{FB}$ taken as the average of all $\Delta(T)$ values except for the data point at $T_c = 50$ K, $\Delta_{FB} = \Delta_{av} \cong 12$ meV (Fig. 7a). Notice that the band is confined to within 10 meV, which is exactly what was found theoretically to corresponds to all 'magic' twist angles (at which the Fermi velocity at the Dirac points becomes zero) [37]. Therefore, the H-C-N fiber appears to be the model system for realizing all possible 'magic' twist angles. The contrast in the $\Delta(T)$ data below and above 50 K might be explained by considering a sort of dynamic equilibrium in the H-C-N fiber.

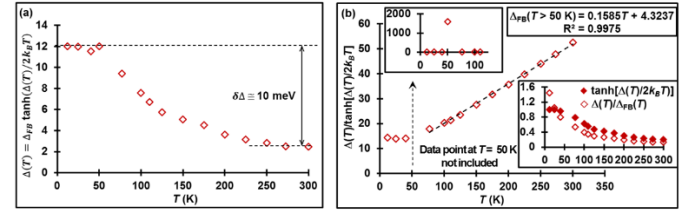

Fig. 7. a) The flat-energy bands $T$-dependence of the gap $\Delta(T)$, where the flat band parameter $\Delta_{FB} \equiv \Delta_{av} \cong 12$ meV. b) $\Delta_{FB}$ is constant below 50 K and linearly increases with $T$ above 50 K. The inset shows that a linear $\Delta_{FB}(T > 50$ K$)$ implies the divergence of $\Delta$ at $T = 50$ K.

This dynamic equilibrium is being determined by the interplay between the two length scales characteristic to the two kinds of order: the smaller periodic crystalline order with the phonons as the major players and the longer quasiperiodic order that is determined by the sequential deposition of different layers and where the $e - e$ correlations can play the major role. We also notice that the gap tends asymptotically to a non-zero constant of about 2 meV at a critical temperature around 300 K. While $\Delta_{FB}(T)$ is $T$-independent below 50 K, the outstanding feature is its linear $T$-dependence for $T > 50$ K (Fig. 7b). The lower inset shows that a linear $\Delta_{FB}(T > 50$ K$)$ implies the divergence of $\Delta(T)$ at $T = 50$ K.

## IV. CONCLUSION

We have found that a H-rich graphite fiber can be an unconventional HTSC. The indirect excitons are the driving force for the boson condensation below $T_c \cong 50$ K. Within the BCS framework, data proves a multi-gap system and show signs for above room-$T$ SC. Analysis of $\Delta(T)$ within the framework of flat-energy bands shows perfectly linear evolution above the temperature for exciton condensation, $T_c \cong 50$ K. Due to the structure of the H-C-N fiber, HTSC and normal nanograins form Josephson normal-SC junctions that can be globally connected by stepping up the applied voltage [2]. The transport properties of Josephson-like normal-SC junctions are determined by the competition between two different coherent reflection processes: a) the standard Andreev reflection at the interface between the SC and the exciton condensate and b) a coherent crossed reflection at the semimetal with the exciton condensate interface that converts electrons from one layer into the other. As a consequence of this competition, $G_{diff}(V)$ has minima at $V$ of the order $\Delta/|e|$, the order parameter for the exciton condensate. Thus, such minima can be seen as a direct hallmark of the existence of a gaped excitonic condensate. The excitonic condensate does carry a net dipolar current, which expression is similar to the one for the supercurrent in a SC, i.e., $\mathbf{J} = |\mathbf{e}|\rho_s(\nabla\Psi - 2|e|\mathbf{A})$, where $\rho_s$ is the density of SC (Cooper) pairs, $\Psi$ is the scalar potential, and $\mathbf{A}$ is the vector potential [38].

We believe that our results and their interpretation are contributing to the mounting evidence of unconventional SC at graphite's interfaces, in particular after the samples were brought in contact with alkanes. There is a mixed structure of SC and normal grains acting as Josephson junctions. Moreover, the existence of flat–energy bands suggests the formation of Majorana edge modes in this FMSC. Others found that H-C systems with even number of C atoms are small SCs [39] and that metallic nanoclusters such of $C_{60}$ can be engineered as HTSC materials [40]. We have previously found evidence of the latter in the H-C-N fiber [2]. Here, we have also estimated for the zero-energy gap a very high value for $T_c \cong 2360$ K ($\approx T_D$) that is also very close to the one found for the SC network in the animal brain [41]. As recently observed [42], typical voltages in a human body vary between 20 and 200 mV, with the average membrane potential of about 70 mV. As already mentioned, a minimum of $G_{diff}(V)$ for $T = 130$ K corresponding to a voltage $V \cong 63$ meV was found. Notice that the voltage values for the human body, corresponding to beyond mid-infrared frequencies 9.7, 96.7 and 33.9 THz, are one order of magnitude smaller than the corresponding frequencies for the interference of asymmetric chiral Andreev edge states found here for the H-C-N fiber. Remarkably, Schrieffer's prediction that a chiral $p$-wave pairing of FM spin fluctuations should lead to super-HTSC [28] suggests a super-high $T_c$ for the human brain.